\newcommand{\be}{\begin{equation}}
\newcommand{\ee}{\end{equation}}

\documentclass{iau-JDSS-hacked}
\usepackage{graphicx,natbib}
\usepackage{natbib}
\def\reff@jnl#1{{\rm#1\/}}
\def\aj{\reff@jnl{AJ}}                  % Astronomical Journal
\def\araa{\reff@jnl{ARA\&A}}            % Annual Review of Astron and Astrophys
\def\apj{\reff@jnl{ApJ}}                % Astrophysical Journal
\def\apjl{\reff@jnl{ApJ}}               % Astrophysical Journal, Letters
\def\apjs{\reff@jnl{ApJS}}              % Astrophysical Journal, Supplement
\def\apss{\reff@jnl{Ap\&SS}}            % Astrophysics and Space Science
\def\aap{\reff@jnl{A\&A}}               % Astronomy and Astrophysics
\def\aapr{\reff@jnl{A\&A~Rev.}}         % Astronomy and Astrophysics Reviews
\def\aaps{\reff@jnl{A\&AS}}             % Astronomy and Astrophysics, Supplement
\def\mnras{\reff@jnl{MNRAS}}            % Monthly Notices of the RAS
\def\prd{\reff@jnl{Phys.Rev.D}}         % Physical Review D
\def\prl{\reff@jnl{Phys.Rev.Lett}}      % Physical Review Letters
\def\pasp{\reff@jnl{PASP}}              % Publications of the ASP
\def\pasj{\reff@jnl{PASJ}}              % Publications of the ASJ
\def\nat{\reff@jnl{Nature}}             % Nature 

\title[] %% give here short title %%
{Tensions between the Early and the Late Universe}

\author[] %% give here short author list %%
{Verde, Licia$^{1}$, Treu, Tommaso$^{2}$, Riess, Adam G.$^{3,4}$}

\affiliation{
  $^{1}$ ICREA \& ICC UB, University of Barcelona, Marti i Franques 1, Barcelona, 08028, Spain., \newline  
  $^{2}$ University of California Los Angeles,
  $^{3}$ Johns Hopkins University, 
  $^{4}$ Space Telescope Science Institute }
 
\pubyear{2012}
\volume{Volume 1}  %% insert here IAU Highlights of Astronomy Volume Number
\pagerange{1-2}
\date{?? and in revised form ??}
\jname{Tensions between the Early and the Late Universe}
\editors{Verde, L., Treu, T., Riess, A.G.}

\begin{document}

\maketitle

\begin{abstract}
The standard cosmological model successfully describes many observations from widely different epochs of the Universe, from primordial nucleosynthesis to the accelerating expansion of the present day. However, as the basic cosmological parameters of the model are being determined with increasing and unprecedented precision, it is not guaranteed that the same model will fit more precise observations from widely different cosmic epochs. Discrepancies developing between observations at early and late cosmological time may require an expansion of the standard model, and may lead to the discovery of new physics.  The workshop ``Tensions between the Early and the Late Universe'' was held at the Kavli
Institute for Theoretical Physics on July 15-17 2019 \footnote{More details of the
workshop (including on-line presentations) are given at the website:\\
\indent\indent\indent \texttt{https://www.kitp.ucsb.edu/activities/enervac-c19}} to evaluate increasing evidence for these discrepancies, primarily in the value of the Hubble constant as well as ideas recently proposed to explain this tension.  
Multiple new observational results for the Hubble constant were presented in the time frame of the workshop using different probes: Cepheids, strong lensing time delays, tip of the red giant branch (TRGB), megamasers, Oxygen-rich Miras and surface brightness fluctuations (SBF) resulting in a set of six new ones in the last several months.  Here we present the summary plot of the meeting that shows combining any three independent approaches to measure $H_0$ in the late universe yields tension with the early Universe values between $4.0\sigma$ and $5.8\sigma$.   This shows that the discrepancy does not appear to be dependent on the use of any one method, team, or source. Theoretical ideas to explain the discrepancy focused on new physics in the decade of expansion preceding recombination as the most plausible. This is a brief summary of the workshop. 


\end{abstract}

\begin{figure}[h!]
\begin{center}
{\includegraphics[width=5.0in]{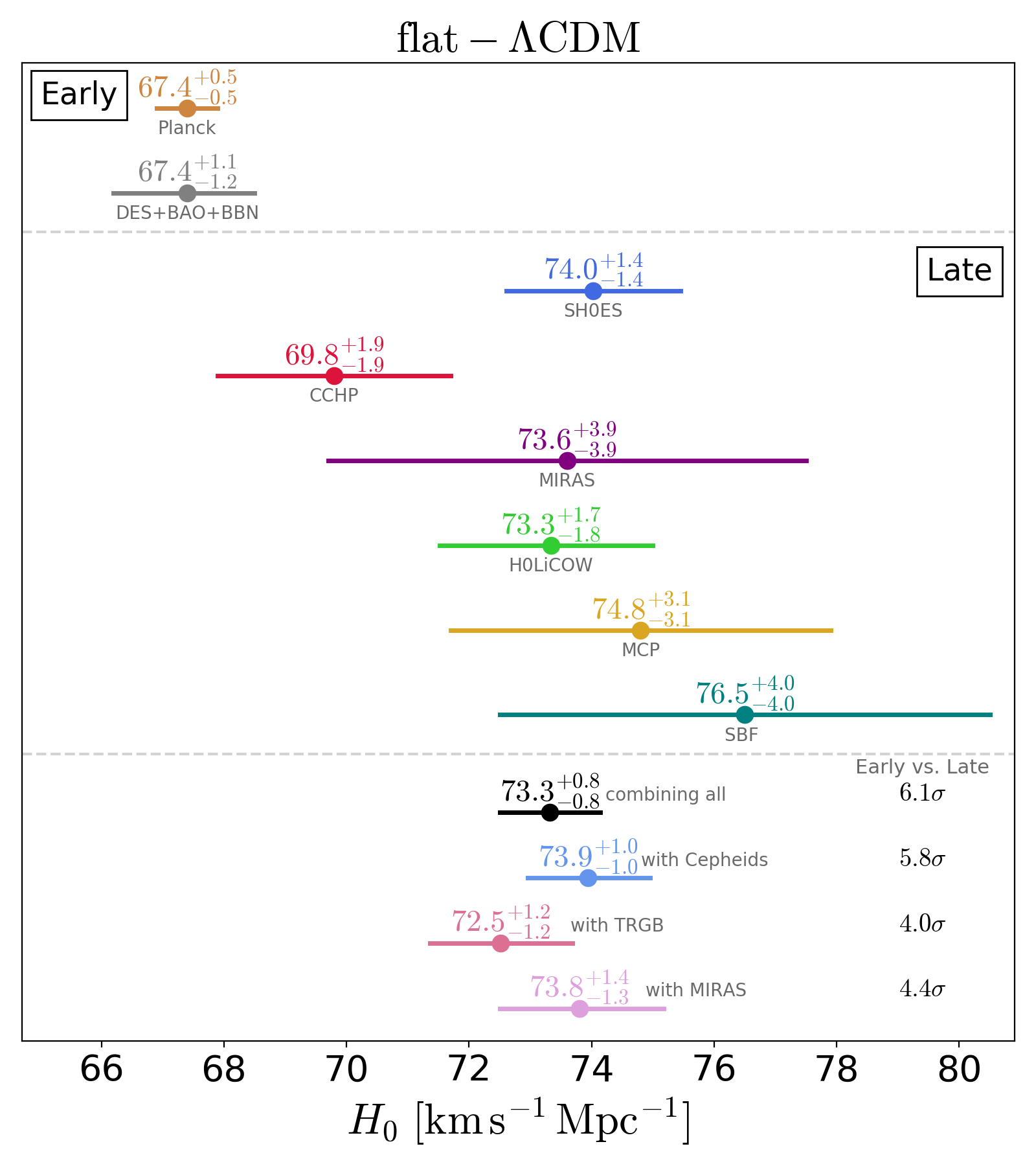}\hskip 0.1in}
\caption{\label{fig:w} Compilation of Hubble Constant predictions and measurements taken from the recent literature and presented or discussed at the meeting. Two independent predictions based on early-Universe data \citep{Planck,Abbott18} are shown at the top left (more utilizing other CMB experiments have been presented with similar findings), while the middle panel shows late Universe measurements. The bottom panel shows combinations of the late-Universe measurements and lists the tension with the early-Universe predictions. We stress that the three variants of the local distance ladder method (SHOES=Cepheids; CCHP=TRGB; MIRAS) share some Ia calibrators and cannot be considered as statistically independent. Likewise the SBF method is calibrated based on Cepheids or TRGB and thus it cannot be considered as fully independent of the local distance ladder method. Thus the ``combining all" value should be taken for illustration only, since its derivation neglects covariance between the data. The three combinations based on Cepheids, TRGB, Miras are based on statistically independent datasets and therefore the significance of their discrepancy with the early universe prediction is correct - even though of course separating the probes gives up some precision. A fair summary is that the difference is more than 4 $\sigma$, less than 6 $\sigma$, while robust to exclusion of any one method, team or source.  Figure courtesy of Vivien Bonvin.}
\end{center}
\end{figure}

\noindent \textbf{1. Introduction: This Moment in Cosmology}\\ 

Nearly a century of cosmological research has led us to a Standard Model of Cosmology, the  $\Lambda$CDM, with its six free parameters and several ansatz.  This model is dominated by dark components (energy and matter) with still uncertain physics.  With some simplifying assumptions about the uncertain bits, the pre-recombination version of the model (63\% dark matter, 15\% photons, 10\% neutrinos and 12\% atoms) is used to predict the physical size of density fluctuations in the plasma, i.e., the sound horizon and its overtones as well as the primordial baryon density.  By comparing the fluctuation spectrum predicted by the model to the angular spectrum observed by the  Cosmic Microwave Background (CMB), the six free parameters are set and the {\it ansatzes} are tested.  An alternative to the use of the CMB for setting the sound horizon may be derived from relating measurements of the primordial deuterium abundance to the predicted baryon density.  The evolving form of the model (68\% dark energy, 27\% dark matter and 5\% atoms) is then used to predict the expansion history of the Universe from $z=1000$ to $z=0$.  Un-calibrated high-redshift type Ia supernovae (SNe) and Baryon Acoustic Oscillations (BAO) provide ``guard rails'' between $z\sim 2$ and 0; they do not tell us if we are on the ``right road'' but they make sure we do not miss the curves in the model's road (i.e., cosmic acceleration consistent with $w=-1$) along the way.  The model calibrated on early Universe observations  predicts  the present-day value of several cosmological parameters, some of which,  can be empirically measured  locally (i.e. $z<1$) with little or no model dependence.   In particular, the model calibrated with Planck data predicts the Hubble constant, today's expansion rate, to a remarkable $1$\% precision, $67.4$ $\pm$ $0.5$ km s$^{-1}$ Mpc $^{-1}$.  Is this whole story right? 

The simplest test of this paradigm, from end to end, is to compare the absolute scale provided through the application of early Universe physics (i.e., the physical size of the sound horizon used to interpret the CMB and BAO) to the absolute scale measured in the Hubble constant in the local, late-time Universe.  Because the Universe has only one true scale and in light of the uncertain physics of the dark sector, comparing the two calibrated at opposite ends is natural and potentially insightful.  (To determine whether a measurement is truly derived from the ``Early'' or ``Late'' Universe it is necessary to trace back its chain of calibration but a useful check is to determine whether or not it depends on, for example, the number of neutrinos assumed in the Standard Model.)   In July 2019, 108 attendees of this workshop gathered to consider growing tensions between the early Universe predictions and the late Universe measurements and how they might be explained.

\medskip
\noindent \textbf{2. The Early Universe Measurements and Predictions for H$_0$}

The early Universe probes were discussed at length. The two key questions were: {\it i)} what kind of cross-checks can be used to identify unknown systematic errors that may affect the predictions for H$_0$? {\it ii)} is there any hint of tension in early-Universe data that may perhaps reveal systematic errors or shortcomings of the standard six-parameter model?

Several talks addressed the first question. In addition to the well known small difference between the inference of H$_0$ from low and high angular resolution Planck and WMAP data, all the early Universe data seem to be consistently predicting a low value of H$_0$ . ACT and SPT are in agreement with Planck, and \emph {any} CMB data used to calibrate the sound horizon and subsequently baryonic acoustic oscillations leads to low $H_0$ of $\sim$ 67-68.5 km s$^{-1}$ Mpc $^{-1}$, even without Planck. A completely independent value and statistically consistent value of $H_0$ can be obtained by using light element abundances to calibrate the sound horizon, BAO and subsequently other lower redshift probes. 

As far as the second question is concerned, some curiosities among high-redshift probes at the level of $\sim$2$\sigma$ were identified. The most compelling ones appear to be the departure from unity of the nuisance parameter A$_{\rm lens}$, used to match CMB anisotropies and CMB-lensing data. If confirmed, this departure from unity represents evidence that something is not well understood in the relation between CMB anisotropies and the growth of structure, and thus could perhaps be a hint towards new physics. The other $2\sigma$ curiosities that were discussed were: {\it i)} the tension between the S$_8$-$\Omega_{\rm m}$ contours inferred from the CMB and those inferred by cosmic shear data; {\it ii)} the tension between the cosmological parameters inferred from  the BAO signal in galaxies at $z<1$ and those of Ly$\alpha$ at higher redshifts and {\it iii)} drifts of the model parameters with the CMB fluctuation scale used to determine the model. The statistical errors of these methods are expected to shrink in the next few years, and will reveal whether the tension is a statistical fluke of the kind that one may expect when considering of order dozens of true and nuisance parameters, or whether it is indicative of some yet-to-be discovered systematic or new physics.  Nevertheless, many wondered if a solution to the Late vs Early Universe discrepancy may be more credible if it also ameliorated one or more of these tensions.

\medskip
\noindent \textbf{3. The Late Universe}

A summary talk from the SH0ES Team presented the status of a 15 year effort to build a consistently measured distance ladder using geometric distances to calibrate Cepheids, followed by 19 hosts to both SNe Ia and Cepheids, followed by hundreds of type Ia supernova in the Hubble flow.  Highlights included the use of near-infrared {\it HST} photometry for all Cepheid data and five independent sources of geometric distances including 3
three types of Milky Way parallaxes, LMC detached eclipsing binaries and the masers in NGC 4258.  The result was $H_0$=74.0 $\pm 1.4$ km s$^{-1}$ Mpc $^{-1}$ \citep{R19}.  

A summary talk by the HoliCOW Team described new results from strong lensing time delays between multiple images of background quasars.  Six such systems have now been measured leading to $H_0$=73.3$^{+1.7}_{-1.8}$ km s$^{-1}$ Mpc $^{-1}$ \citep{Holicow}.  Robustness was demonstrated by the use of lens quads, doubles, long and short time delays, imaging from {\it HST} or {\it Keck AO} \citep{Chen} and different mass modelling softwares \citep{Birrer}.  Because each lens model must be constructed individually for each system, the lensing analysis was carried out blindly to avoid experimenter bias in the model construction and it is completely independent of the local distance ladder method. The SHOES and H0LICOW results were the only known shortly before the meeting and together provide a 5.3$\sigma$ difference with the Early Universe.

New results were presented from the Megamaser Cosmology Project \citep[MCP][]{Reid} which uses VLBI observations of water masers in circum-nuclear orbits around supermassive black holes to measure geometric distances.   A much improved measurement of the distance to the nearby NGC 4258 (similar distance but full error reduced from 2.6\% to 1.5\%) was presented.  In addition, a longer time span of VLBI measurements and improved analysis of the distances to four other masers in the Hubble flow, UGC 3789, CGCG 074-064, NGC 5765b and NGC 6264 were presented which do not require a distance ladder and together yielded $H_0$=74.8 $\pm 3.1$ km s$^{-1}$ Mpc $^{-1}$.  

New results were presented by the CCHP collaboration which used Tip of the Red Giant Branch (TRGB) measurements in the LMC to calibrate 18 SN Ia (across 14 hosts) in lieu of Cepheids to connect the distance ladder \citep{TRGB}.  Pros and cons of TRGB and Cepheids were extensively discussed across three talks.  TRGB was recognized as a valuable independent tool, well understood from first principles that can be observed on simple backgrounds. The CCHP result of $H_0$=69.8 $\pm 1.9$ km s$^{-1}$ Mpc $^{-1}$ based on a new calibration of TRGB in the LMC was presented \citep{TRGB}.  The source of the reduction in this value over the prior TRGB result from \cite{JL17} was extensively discussed and identified as traceable to a 0.08 mag increase in the estimate of the LMC TRGB $I$-band luminosity of which 0.06 $\pm$ 0.02 mag was attributed to a different method for estimating TRGB extinction in the LMC which would find it is 3 $\sigma$ greater than the values given by the OGLE reddening maps (the TRGB color method yields $A_I=0.16 \pm 0.02$ vs the use of reddening maps used by \cite{JL17} gave $A_I=0.10$).  

New results were presented measuring IR Surface Brightness Fluctuations to two new sets of {\it HST} imaging of Early type hosts from the {\emph massive} sample and a sample of SN Ia hosts raising the total sample from $N=15$ to $N=54$ galaxies out to 100 Mpc \citep{SBF}.  The result was $H_0$=76.5 $\pm 4.0$ km s$^{-1}$ Mpc $^{-1}$.  Important features of this measurement are that it is fully independent of the use of SNe Ia (which other distance ladders use) and was shown to vary within the error by altering the source of the calibration of the SBF luminosity from Cepheids, TRGB and stellar population models.

New results were presented using oxygen-rich variable stars at the tip of AGB stars (i.e., Miras) observed in the NIR in lieu of TRGB or Cepheids to connect the distance ladder.  Pros of Miras are that they are brighter than TRGB and offer an older population than Cepheids, present in ellipticals and halos of spirals.  Cons include the potential confusion with C-rich Miras and those undergoing hot bottom burning, but this issue can be mitigated by using the period range $P < 400$ days.   Results from the SH0ES Team (Huang et al. 2019, 2018) were presented using Mira measurements in $F160W$ from HST in NGC 4258 and in the halo of the farthest Mira host and first SN Ia host to date, NGC 1559 at a distance near 20 Mpc.  A distance ladder from Miras to connect the geometric distances to the LMC and NGC 4258 and then to calibrate SNe Ia yielded $H_0$=73.6 $\pm 3.9$ km s$^{-1}$ Mpc $^{-1}$ with the error dominated by the 5\% uncertainty in the calibrated luminosity of a single SN Ia, but observations for 3 other SNe Ia were reported to be in progress.  

With two recent results and four new results, a summary plot was made (see Figure 1) with approximate data combinations yielding a 4 to 6 $\sigma$ discrepancy with the early Universe result.  The use of all data yielded a 6 $\sigma$ difference and $H_0=$ 73.3 $\pm 0.8$ km s$^{-1}$ Mpc $^{-1}$.  However, there is some overlap in data between the 3 ladders (Cepheids, TRGB and Miras) which connect geometrical distances to SNe Ia, and therefore they cannot be simply averaged without accounting for their covariance.  In order to give the reader a set of truly independent datasets and a feel for the impact of removing some experiments, combinations using only one of these at a time (i.e., eliminating the other two) are also shown.  The combination including Cepheids yields $H_0=$ 73.9 $\pm 1.0$ km s$^{-1}$ Mpc $^{-1}$ and a 5.8$\sigma$ difference.  Miras yield a similar combined result but with lower (4.4 $\sigma$) significance due to the small number of Mira/SN hosts.  TRGB (sans Cepheids or Miras) combined with the other measurements gives $H_0=$ 72.5 $\pm 1.2$ km s$^{-1}$ Mpc $^{-1}$ and a $4 \sigma$ difference.  A more careful combination of all the data accounting for all covariance would thus be expected to give a result between 4-6 $\sigma$.  The use of all methods which do not use SNe Ia (Masers, SBF and strong lensing) also yield more than a 4 $\sigma$ discrepancy with respect to the early Universe.  

Extensive use was made of the new 1\% measurement of the distance to the LMC from 20 detached eclipsing binaries \citep{Pietrzynski} and there was some discussion on how to extend the method to other hosts.

Three talks discussed the present status of Gaia DR2 parallaxes and approaches to calibrate their zeropoint uncertainty whose value was seen to increase with brightness and redness (away from the faint, blue quasar sample used to provide the initial value).  There was much optimism that this issue would be further settled by the time of DR3.  

Many talks and ensuing Q\&A sessions discussed future prospects of each method and the potential to reach sub-percent precision on H$_0$ from each individual method. 

For the traditional Cepheid-based local distance ladder, future precision was expected to reach 1.3\% with caution expressed that a notional goal of 1.0\% would be hard to reach.  Improvements in the TRGB and Miras-based distance ladders are also expected with the launch of the James Webb Space Telescope and the upcoming Gaia data releases. The precision of time-delay cosmography is currently limited by sample size. With the recent explosion of discovery of quadruply imaged quasars in wide field imaging surveys \citep{Shajib} and the discovery of the first lensed supernovae \citep{Kelly} it is clear that the sample size is not a limitation anymore. The current limitation is follow-up and the scientist-time required for high-precision lens models. No source of systematic has been discovered so far that would prevent reaching sub-percent precision from this method, even though of course this has to be demonstrated in practice and through data challenges. The precision of the Megamaser Cosmology Project is also limited by sample size and although the number of masers at the right distance is finite, there appears to be room for further improvements in precision. Likewise, there is room for further improvement in precision of the SBF method, even though it is unclear whether either method can ultimately reach sup-percent precision. 

A (still blind) time delay for the multiply imaged SN Refsdal was presented en route to a determination of $H_0$ to an anticipated statistical precision of $\sim 7$\% \citep{Grillo}. There was much excitement also about the prospects of gravitational waves and standard sirens to contribute to the conversation. Even though the current sample and corresponding precision is not competitive with other methods \citep{Abbott17}, based on the forecasts shown the method will soon be another powerful and independent tool at our disposal.

Completely independent of the distance determination methods is the cosmic chronometers method discussed at this meeting. In contrast to everything that we have summarized so far, the method has the advantage of measuring age (as opposed to distance) and thus it is perhaps the only one that can directly answer the existential question "how old is the universe?" with minimal cosmological assumptions. \cite{Clocks19}  reports $t_U=13.2\pm 0.44$Gyr from 22 Globular clusters \citep{GCages}. In a $\Lambda$CDM model the ages of these objects implies  $H_0=71.0\pm 2.8$ km s$^{-1}$Mpc$^{-1}$. On the other hand, relative ages of suitably selected old passively evolving elliptical galaxies (e.g., \cite{chronometers1, chronometers2} and refs therein) yield an estimate of $H(z)$ with most statistical power at $z=0.43$, $H(z=0.43)=91.8\pm 5.3$ km s$^{-1}$Mpc$^{-1}$. The next step in these measurements should involve a better characterization of model uncertainties, perhaps through a survey of the model space.  Such efforts are underway.  As the statistical precision of the method progresses, comparison with other probes and internal consistency checks will tell what is the systematic noise floor related to our ability to determine ages of stellar populations.

The initially reported tension between the uncalibrated BAO distances at $z_{\rm eff}=2.3$ via Lyman$\alpha$ forest  and the  one measured at $z=0.75$ has decreased to 1.7$\sigma$  with the latest data, and therefore not significant enough to be considered a true tension\citep{Addison}.
The tension in  the $\sigma_8$ parameter (or equivalently   $S_8=\sigma_8\sqrt{\Omega_m/0.3}$) as inferred by Planck  and measured by weak gravitational lensing surveys is below the $3\sigma$ level, although the exact significance depends on the lensing data set chosen and assumptions made in the analysis. While investigation is important, this ``tension" is not as dramatic as in the $H_0$ parameter. Two considerations are in order, however. {\it i)} This may be related to the Planck  internal consistency test offered by the parameter $A_{\rm lens}$. This parametrizes the gravitational lensing amplitude in CMB data and of course depends very closely on the amplitude of perturbations (i.e., $\sigma_8$). When inferred from the smoothing of the high $\ell$ angular temperature power spectrum peaks, its value is $_{>}^{\sim} 2 \sigma$ away from that inferred from CMB lensing signal.   {\it ii)} Any new physics introduced to explain the $H_0$ discrepancy should not make the $\sigma_8$ tension significantly worse.

\medskip
\noindent \textbf{4. Ideas to reconcile the two}\\

This leaves us with the question: how can the $H_0$ discrepancy be solved?

The most skeptical approach is to invoke systematic errors in the data. However given the size of the discrepancy and the independence of routes seeing it, a single systematic error cannot be the explanation. It also should be said that following this approach too strongly is to lose the ability to make fresh discoveries.
\footnote{A more formal approach to invoking multiple, unknown systematic errors follows the {\it BACCUS} approach \citep{baccus}. This was not presented at the meeting, but since then, one of us (LV) has experimented with this approach using shifts only, to combine the late-Universe $H_0$ determinations marginalizing over (unknown) possible systematic shifts for each measurement.  The  shifts are assumed to be drawn from the same prior distribution: a Gaussian with width $\sigma_a$, which gets marginalized over with a uniform hyperprior $-10<\sigma_a<10$ km/s/Mpc.  As expected, {\it BACCUS} combination widens the tails of the  resulting posterior distribution compared to conventional (Gaussian) combination, but does not single out any local Universe measure as a-typically shifted compared to the others.
The combination of all independent measurements (SH0ES, HoLiCOW, SBF, MPC, TRGB, and GC ages) yields a grand average of $H_0=72.7\pm 1.2(\pm 2.9)$ km s$^{-1}$ Mpc$^{-1}$ at 68(95)\% confidence levels, nominally still 4 $\sigma$ from the Early Universe value, but the widening of the posterior tails implies that this is $2.6 (2.8) \sigma$ or 99.1\%(99.5\%) CL away from a $H_0$ value of 68 (67.4) km s$^{-1}$ Mpc$^{-1}$.  However, it should be noted that the conventional high bar of $\sim$ 5$\sigma$ (one in a million experiments) as a discovery threshold already invokes the pessimistic approach of expecting the presence of unknown systematic errors by requiring margin.  The {\it BACCUS} approach provides an alternative method to invoke this pessimist's prior so a $>$ 99\% CL after employing it is quite significant.  A skeptic could require a high discovery threshold {\it or} use the {\it BACCUS} approach but presumably not both without a strong prior against the possibility of new physics.}

After a thorough re-analysis and cross checks of multiple CMB observations (Planck, SPT, ACT etc.), it is clear that systematic errors in CMB data cannot alone explain the tension.

Moreover,  a suite of  low redshift, different, truly independent measurements,  affected by completely different possible systematics, agree with each other; it seems improbable that completely independent systematic errors affect all these measurements by shifting them all by about the same amount and in the same direction.   

An obvious but important caveat is that, if this tension is an indication of new physics beyond $\Lambda$CDM, the new model should not do worse than standard $\Lambda$CDM in describing all other cosmological observations!

For example there is not much freedom in changing the expansion history from that of a standard $\Lambda$CDM model below $z\sim 2$: the guardrails offered by SNe and BAO do not allow this.
Moreover, model changes away from $\Lambda$CDM are tightly constrained by CMB data \citep{Knox1, Knox2}.

The Early-Universe $H_0$ determination relies on  angular scales such as  sound horizon (at radiation drag) and  matter-radiation equality. These angular scales are extremely well determined by CMB data, but they depend on a ratio of two qualitatively different quantities: the physical scale (which depends on early-time physics and background parameters --such as  physical densities of  matter $\omega_m$, of baryons  $\omega_b$  etc.--)  and the angular size distance to the CMB (which depends on $H_0$ as well as other background parameters). To keep the angular scales fixed while increasing $H_0$, both the physical scales and the distance must decrease.  
To reconcile the $H_0$ values the CMB-inferred sound horizon  at radiation drag should be lowered by $\sim 7$\%, but any  new physics  should only affect the decade of expansion before recombination; changes from $\Lambda$CDM in other windows would worsen the fit to existing data. In particular  any change in background parameters (physical densities) should be mostly  via $H_0$ and not via  the density parameters themselves. Few examples to achieve this were presented. 
One possibility is a scalar field  acting as an early dark energy component \citep{Poulin}.  The dynamics of the field are constructed so that the energy density of this early dark energy component is relevant only over a narrow epoch in the expansion history of the universe: after matter-radiation equality but before recombination. Such model  yields a higher value for the  CMB-inferred $H_0$, greatly alleviating  the tension and, notably, preserving a good fit to all relevant observations (CMB, BAO, SNe etc.). The epoch immediately preceding recombination is favored because it is the time when the bulk of the sound horizon accrues.  
More data, especially CMB polarization measurements and/or very low $\ell$ measurements at greater precision, however are needed to test other predictions of the model and to determine whether an additional early component (and the extra parameters that this model introduces) is actually favoured over a $\Lambda$CDM model.

Another family of possibilities was presented that extends instead the radiation sector of the early Universe physics. A solution invoking  extra free streaming neutrinos is penalized by a worse fit to high $\ell$ CMB angular power spectra where the specific gravitational coupling of free streaming neutrinos leaves its  signature. However, this behavior may be offset by invoking  neutrinos with self-interactions so that neutrinos do not free stream but rather behave like tightly coupled radiation \citep{Cyr-Racine1}. A model which allows neutrinos self interactions and additional neutrinos species, if compared to the standard data set combination of CMB and BAO data, produce an allowed region in parameter space  which is  characterized  by  high $H_0\simeq 72$km s$^{-1}$ Mpc$^{1-}$,  an extra effective neutrino specie $N_{\rm eff}\sim 4$ but very strong coupling.  Such a solution predicts specific signatures in the matter power spectrum that may be sought. However, is difficult to achieve the strong interaction that this solution finds  from a particle model-building perspective, while still evading other constraints on neutrino physics. 
Also in this case future data may either find  signatures  of the other predictions of this model or rule it out.

Similarly, models with an extra scalar field that provides energy injection localized  around matter-radiation equality\citep{Cyr-Racine2} also improves the fit to late-time $H_0$ determinations  (still being consistent with BAO data) at little or even no cost to the fit to CMB data.  While none of these models may appear natural, similar or even greater tuning may already be required to explain the two other accepted episodes of dark energy, raising the question, is twice okay but three too many?

As precision increases one may wonder if cracks may be appearing in the $\Lambda$CDM model.
It may be possible to find models that are radically different from $\Lambda$CDM and still provide good fits to the data including fixing the "cracks". These are likely to have their own specific signatures be it in other cosmological observable or particle physics experiments which will be crucial to make further progress.  
To summarize, the final speaker (Cyr-Racine) concluded,
``We have yet to identify a complete solution that is palatable to both cosmologists and particle physicists, but have important clues about what a successful model would look like".

\medskip
\noindent \textbf{5. Concluding Remarks} 

During the last talk, prior to hearing about possible theoretical solutions to the tension, a draft version of Figure~1 was shown to the audience and the audience was asked to vote on its perception of the significance of the tension on the following scale: 2$\sigma$=curiosity; 3$\sigma$=tension; 4$\sigma$=discrepancy or problem; 5$\sigma$=crisis. This clearly tongue-in-cheek experiment was carried out to evaluate what kind of Bayesian prior the attendees applied to the evidence. 
Most attendees voted for a Hubble constant ``problem" with tails in favor of both ``tension" and ``crisis" and with no support for something less.  Therefore it appears that the issue is serious, not only taking the uncertainties at face value, but also in the eyes of the community as represented by the KITP workshop participants.  

It was also clear that a great deal of progress has been recently made using new methods to tackle previously difficult measurements.  Little can be learned by simply invoking a checkered past without critical study of the developments in these new measurements.

Finally, going forward, the resolution to the ``problem" will likely require a coordinated effort from the side of theory, interpretation, data analysis and observations. To streamline the interaction between these different communities and more transparent transfer of information, participants advocated adopting  the following  best practices:

\begin{itemize}
\item[*]Model assumptions: one should always make clear where the cosmology dependence enters in a measurement or interpretation.
\item[*] Reproducibility:  one should release data and non-trivial software publicly (see for example CLASS /CAMB codes for cosmology, whenever possible it would be useful to provide any new tools as public plug-ins for those).  Requests for data from published results should be fulfilled promptly.
\item[*]Data transparency: One should release more low-level data products where the least (cosmological) assumptions have been made
\item[*]Blinding:  Blind analysis should be done whenever possible but especially if analysis choices must be made.  Alternatively, the impact of choices should be clearly presented in variants of the primary analysis.
\item[*] When combining CMB data and late-time data for models that address the Hubble problem, one should also present results from CMB data alone. In a successful model, the addition of the low-redshift data should not degrade the fit to the CMB data.  
\item[*] Data challenges: whenever possible, organize mock data challenges designed to blindly test the accuracy and precision of the methods and hypotheses adopted by the community.
\end{itemize}
\vspace*{0.2cm}

We thank KITP for hosting and supporting this workshop. We thank all the participants for an extremely lively and productive workshop. We are also grateful to those who contributed to the scientific results presented at the meeting and could not attend in person. Finally, we are especially grateful to Vivien Bonvin and Anowar Shajib for updating and producing versions of Figure~1 during the meeting, keeping up with the fast pace of presentation of new results.

\bibliography{references}

\begin{thebibliography}{}
%\bibliography{H0Communique.bib}

\bibitem[Abbott et al.(2017)]{Abbott17} Abbott, B.~P., Abbott, R., Abbott, T.~D., et al.\ 2017, \nat, 551, 85

\bibitem[Abbott et al.(2018)]{Abbott18} Abbott, T.~M.~C., Abdalla, F.~B., Annis, J., et al.\ 2018, MNRAS, 480, 3879

\bibitem[Addison et al.(2018)]{Addison} Addison, G.~E., Watts, D.~J., Bennett, C.~L., et al.\ 2018, ApJ, 853, 119

\bibitem[Agrawal et al.(2019)]{Cyr-Racine2} Agrawal, P., Cyr-Racine, F.-Y., Pinner, D., et al.\ 2019, arXiv e-prints, arXiv:1904.01016

\bibitem[Knox et al.(2020)]{Knox2} Aylor, K., Joy, M., Knox, L., et al. \ 2019, in prep.

\bibitem[Aylor et al.(2019)]{Knox1} Aylor, K., Joy, M., Knox, L., et al.\ 2019, ApJ, 874, 4

\bibitem[Bernal, \& Peacock(2018)]{baccus} Bernal, J.~L., \& Peacock, J.~A.\ 2018, JCAP, 2018, 002

\bibitem[Birrer et al.(2019)]{Birrer} Birrer, S., Treu, T., Rusu, C.~E., et al.\ 2019, \mnras, 484, 4726

\bibitem[Chen et al.(2019)]{Chen} Chen, G.~C.-F., Fassnacht, C.~D., Suyu, S.~H., et al.\ 2019, arXiv e-prints, arXiv:1907.02533

\bibitem[Freedman et al.(2019)]{TRGB} Freedman, W.~L., Madore, B.~F., Hatt, D., et al.\ 2019, arXiv e-prints, arXiv:1907.05922

\bibitem[Grillo et al.(2018)]{Grillo} Grillo, C., Rosati, P., Suyu, S.~H., et al.\ 2018, \apj, 860, 94

\bibitem[Huang et al.(2018)]{Huang} Huang, C.~D., Riess, A.~G., Hoffmann, S.~L., et al.\ 2018, \apj, 857, 67

\bibitem[Jang and  Lee(2017)]{JL17} Jang, I.~S., \& Lee, M.~G.\ 2017, ApJ, 835, 28

\bibitem[Jimenez et al. (2019)]{Clocks19} Jimenez R., Cimatti A., Verde L., Moresco M., Wandelt B., 2019, JCAP, 2019, 043

\bibitem[Kelly et al.(2015)]{Kelly} Kelly, P.~L., Rodney, S.~A., Treu, T., et al.\ 2015, Science, 347, 1123

\bibitem[Kreisch et al.(2019)]{Cyr-Racine1} Kreisch, C.~D., Cyr-Racine, F.-Y., \& Dor{\'e}, O.\ 2019, arXiv e-prints, arXiv:1902.00534


\bibitem[Moresco et al.(2018)]{chronometers2} Moresco, M., Jimenez, R., Verde, L., et al.\ 2018, ApJ, 868, 84

\bibitem[Moresco et al.(2016)]{chronometers1} Moresco, M., Pozzetti, L., Cimatti, A., et al.\ 2016, JCAP, 2016, 014

\bibitem[O{\textquoteright}Malley et al.(2017)]{GCages} O{\textquoteright}Malley, E.~M., Gilligan, C., \& Chaboyer, B.\ 2017, ApJ, 838, 162

\bibitem[Pietrzy{\'n}ski et al.(2019)]{Pietrzynski} Pietrzy{\'n}ski, G., Graczyk, D., Gallenne, A., et al.\ 2019, \nat, 567, 200

\bibitem[Planck Collaboration et al.(2018)]{Planck} Planck Collaboration, Aghanim, N., Akrami, Y., et al.\ 2018, arXiv e-prints, arXiv:1807.06209

\bibitem[Potter et al.(2018)]{SBF} Potter, C., Jensen, J.~B., Blakeslee, J., et al.\ 2018, American Astronomical Society Meeting Abstracts \#232 232, 319.02

\bibitem[Poulin et al.(2019)]{Poulin} Poulin, V., Smith, T.~L., Karwal, T., et al.\ 2019, PRL, 122, 221301

\bibitem[Reid et al.(2009)]{Reid} Reid, M.~J., Braatz, J.~A., Condon, J.~J., et al.\ 2009, \apj, 695, 287

\bibitem[Riess et al.(2019)]{R19} Riess, A.~G., Casertano, S., Yuan, W., et al.\ 2019, APJ, 876, 85

\bibitem[Shajib et al.(2019)]{Shajib} Shajib, A.~J., Birrer, S., Treu, T., et al.\ 2019, \mnras, 483, 5649

\bibitem[Wong et al.(2019)]{Holicow} Wong, K.~C., Suyu, S.~H., Chen, G.~C.-F., et al.\ 2019, arXiv e-prints, arXiv:1907.04869

%%\bibliography{references}
%
%\begin{thebibliography}{}
%
%\bibitem[Bolton et al.(2008)]{Bol++08} Bolton, A.~S. et al.\ 2008,
%\textit{ApJ}, 684, 248
%
%\bibitem[Bolton et al.(2007)]{Bol++07} Bolton, A.~S. et al.\ 2007 
%\textit{ApJL}, 665, L105
%
%\bibitem[Djorgovski \& Davis(1987)]{D+D87} Djorgovski, S., \& Davis,
%M.\ 1987, \textit{ApJ}, 313, 59
%
%\bibitem[Ciotti et al.(1996)]{CLR96} Ciotti, L., Lanzoni, B., \&
%Renzini, A.\ 1996, \textit{MNRAS}, 282, 1
%
%\bibitem[Dressler et al.(1987)]{Dre++87} Dressler, A., et al.  1987,
%\textit{ApJ}, 313, 42
%
%\bibitem[Nipoti et al.(2009)]{Nip++09} Nipoti, C., Treu, T., Auger,
%M.~W., \& Bolton, A.~S.\ 2009, \textit{ApJL}, 706, L86
%
%\bibitem[Treu et al.(2009)]{Tre++09} Treu, T. et al. 2009,
%\textit{ApJ}, submitted, arXiv:0911.3392
%
\end{thebibliography}
\bibliographystyle{apj}

\end{document}